# Effect of ferromagnetic exchange field on band-gap and spin-polarization of graphene on a TMD substrate

Partha Goswami

Physics Department, D.B.College, University of Delhi, Kalkaji, New Delhi-110019, India

**Abstract** We calculate the electronic band dispersion of graphene monolayer on a two dimensional transition metal dichalcogenide substrate (GTMD) (viz., $XY_2$, X = Mo, W; Y = S, Se) around **K** and **K′** points taking into account the interplay of the exchange field due to the ferromagnetic impurities and the substrate induced, sub-lattice-resolved, strongly enhanced intrinsic spin-orbit couplings(SOC). There are extrinsic Rashba spin-orbit coupling(RSOC) and the orbital gap related to the transfer of the electronic charge from graphene to $XY_2$ as well. The former allows for external tuning of the band gap in GTMD and connects the nearest neighbors with spin-flip. On account of the strong SOC, the system acts as a quantum spin Hall insulator. We introduce the exchange field ($M$) in the Hamiltonian to take into account the deposition of Fe atoms on the graphene surface. The cavalcade of the perturbations yield particle-hole symmetric bands with an effective Zeeman field due to the interplay of the substrate induced interactions with RSOC as the prime player. Our graphical analysis with extremely low-lying states strongly suggests the following: The GTMDs like $WY_2$ exhibit band gap narrowing/widening (Moss-Burstein(MB)gap shift)including the spin-polarization inversion(SPI) at finite but low temperature (T ~ 1 K) due to the increase in the exchange field ($M$) at the Dirac point **K**. For graphene on $MoY_2$, on the other hand, the occurrence of the MB-shift and the SPI at higher temperature (T ~ 10 K) take place as $M$ is increased at the Dirac point **K′**. Finally, there is anti-crossing of non-parabolic bands with opposite spins around Dirac points. A direct electric field control of magnetism at the nanoscale is needed here. The magnetic multiferroics, like $BiFeO_3$ (BFO), are useful for this purpose due to the coupling between the magnetic and electric order parameters.

**1. Introduction** The theoretical and the experimental investigations in material science took a completely different turn ever since graphene was isolated and produced as a two-dimensional material in 2004 **[1,2]**. It was found to possess host of unusual properties, such as the high carrier mobility, the low resistivity, the large in-plane stiffness**[3]**, the larger than metal optical absorption at the Plasmon-resonance **[4]**, and so on. Despite these remarkable properties, it has not been possible to fully exploit the graphene's potential due to the difficulty of opening a reasonably sized insulating gap in its band structure. The absence of the gap is owing to graphene's weak spin-orbit coupling. As a result, the attention of the material science community had begun to shift to other two-dimensional (2D) systems such as transition metal dichalcogenide (TMD)**[5,6,7]** (like molybdenum disulfide ($MoS_2$), tungsten diselenide ($WSe_2$),etc.), phosphorene **[8]**, silicene **[9]**, and lately on hexagonal monolayers made up of group IV and VI elements **[10,11,12,13]**, viz. SnS, SnSe, GeS, and GeSe, etc.. The hunt for the new 2D materials is on.

The investigations of the authors in ref.**[6]** showed that a single layer of TMDs is so photo-sensitive that it can capture more than 10% of incoming photons. This process can also be reversed to turn electricity into light. This extra-ordinary ability makes TMDs very promising candidates for applications in communications and quantum cryptography. The silicene (Si monolayer with buckled structure) and phosphorene (phosphorus monolayer with puckered structure), on the other hand, have opened up the possibility of the use of group IV/V based 2D materials for electronics applications**[9,14,15,16,17]**. The silicene allows creation of an electric-field tunable band-gap, but like graphene it is a better conductor of electrons than most TMDs. Moreover, the instability and the reactivity of a monolayer silicene in air is phenomenally high. Thus, it fails act as an appropriate platform for the digital electronics. Interestingly, the phosphorene has an inherent, direct and appreciable band-gap that depends on the number of layers. It is shown to act as a field effect transistor**[17]**. Though it is more stable than silicene, it, however, misses the appropriate platform mark as it also conducts electrons very swiftly. As regards{SnSe, GeS,....}, these semi-conducting materials undergo an indirect-to-direct gap transition by the application of mechanical strain and could be used as LEDs. The suggestion has come forth, side-by-side these new 2D materials, from several workers and collaborators **[18]** to combine them with known 2D materials in such a way that all their different advantages are properly utilized. The appropriate hetero-structure of different layers of two-dimensional materials could have very promising properties and, consequently, much wider applications than previously thought. For example, optically transparent and conductive graphene on very photosensitive TMD (G+TMD) surface could collectively create a very efficient photovoltaic device. In principle, novel, multifunctional devices could be created possibly from every hetero-structure.

The central interaction in topological insulator(TI)– a new phase of matter with their topological nature protected by time reversal symmetry(TRS) [19,20,21,22]– is the strong spin-orbit coupling, where

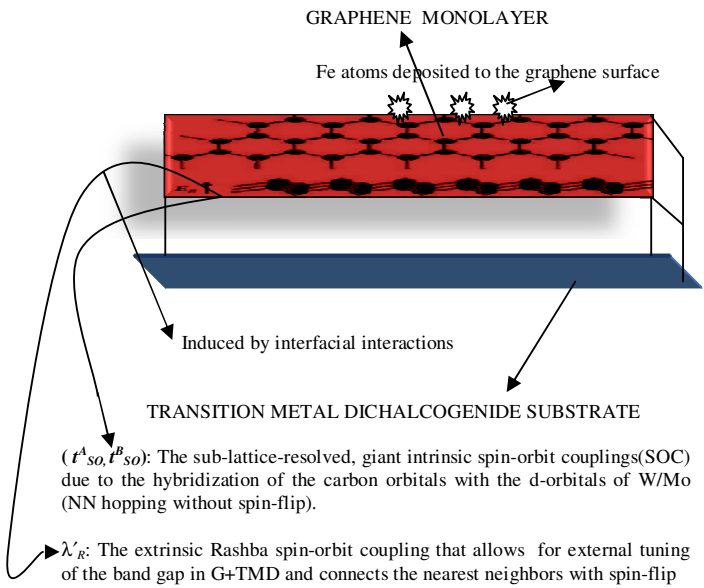

**Schematic Diagram :** Four substrate-induced interactions, viz. $\Delta_{orbital}$ $t^A_{SO}$, $t^B_{SO}$ and $\lambda'_R$

orbital gap related to the transfer of the electronic charge from graphene to TMD

GRAPHENE MONOLAYER

Fe atoms deposited to the graphene surface

Induced by interfacial interactions

TRANSITION METAL DICHALCOGENIDE SUBSTRATE

($t^A_{SO}, t^B_{SO}$): The sub-lattice-resolved, giant intrinsic spin-orbit couplings(SOC) due to the hybridization of the carbon orbitals with the d-orbitals of W/Mo (NN hopping without spin-flip).

$\lambda'_R$: The extrinsic Rashba spin-orbit coupling that allows for external tuning of the band gap in G+TMD and connects the nearest neighbors with spin-flip the quantum spin Hall effect (QSHE)[23] was observed experimentally. This effect allows the spin degrees of freedom to be controlled by electrical means. The non-trivial topological origin is the important feature of the transport properties of TIs which result in dissipation-less charge or spin current carried by edge states with conductivity quantized in units of $e^2/h$. The QSHE state, however, has inherent constraints, such as the requirement of a large magnetic field, difficulty in manipulating the spin degree of freedom by controlling external fields due to spin degeneracy, and so on. In order to avoid such constraints, one may consider the quantum anomalous Hall effect (QAHE)phase[24,25] with broken TRS hosted by TIs. *The quantization of the transverse charge conductivity in a material with intrinsic non-vanishing magnetization* is the essence of the QAHE. This effect was first proposed by F.D.M. Haldane [26] in a two-dimensional (2D) honeycomb lattice with locally non-vanishing magnetic flux but zero in average.

**Figure 1.** A schematic diagram of the graphene on two dimensional TMD. The four substrate-induced interaction terms, $\Delta_{orbital}$ $t^A_{SO}, t^B_{SO}$, and $\lambda'_R$, shown in the figure, are time-reversal invariant and absent by inversion symmetry in isolated pristine, pure graphene sheets.

Kane and Mele [22] first suggested the existence of the quantum spin Hall effect(QSHE) for pure graphene when moderate to large SOC is taken into account. Recently there have been reports of experimental observation of the spin–orbit coupling (SOC) enhancement in graphene through addition of SOC-active impurities[27,28]. Such proximity SOC leads to the spin Hall effect even at room temperature. Theoretically, also the same conclusion was arrived at assuming the typical size of SOC-active impurities is much larger than the lattice spacing, hence suppressing inter-valley scattering [29]. We wish to theoretically show, in this paper, interesting possibilities due to the engineering of the enhanced spin-orbit coupling (SOC) in graphene through interfacial effects via coupling to the suitable substrates, viz. a two dimensional transition metal dichalcogenide (G+TMD); the graphene layer is exchange(M)coupled to the magnetic impurities, such as Fe atoms deposited to the graphene surface as well. The corresponding schematic diagram is shown in Figure 1. The four substrate-induced interaction terms, shown in the figure, are time-reversal invariant and absent by inversion symmetry in isolated pristine, pure graphene monolayer. This leads to : (i) the QSH state for M = 0, (ii) the accessibility of the QAH state when TRS is broken, (iii) the particle-hole symmetric bands with an effective Zeeman field due to the interplay of the substrate induced interactions with RSOC as the prime player,(iv) the anti-crossing of the non-parabolic bands with opposite spins around Dirac points, (v) the band-gap narrowing/widening (for a certain range of the exchange field values) due to the presence of the exchange field, (vi) spin-polarization inversion, and so on. It may be mentioned that it is experimentally established [30,31,36] that a ferro-magnetic topological insulator exhibits the QAH state. The anti-crossing of the non-parabolic bands, on the other hand, have been shown by MacDonald et al.[37] several years ago. Since the the fifth and the sixth findings are novel ones for the G + TMD system, our discussion will be centered around the same. On a quick side note, the present work is motivated by the series of the theoretical investigation on the same system by M. Gmitra et al [32,33,34,35,37]. The experimental finding of a gap of 0.26 eV when graphene is epitaxially grown on the SiC substrate[36] is another motivating result which gives justification of considering a substrate induced interaction and the corresponding gap $\Delta_{orbital}$ . This gap increases as the sample thickness decreases. It has been proposed that the origin of this gap is the breaking of sub-lattice symmetry owing to the graphene-substrate interaction. A direct, functional electric field control of magnetism at the nano-scale is needed for the effective demonstration of our result. The magnetic multi-ferroics, like $BiFeO_3$ (BFO) have piqued the interest of the researchers world-wide with the promise of the coupling between the magnetic and electric order parameters.

The paper is organized as follows: In Sec. II, a brief outline of the low-energy model of the graphene (GTMDC) monolayer on two dimensional transition metal dichalcogenides is given and the single-particle excitation spectrum is obtained. The spectrum is obtained from a quartic, involving all the substrate induced perturbations, and not an approximate bi-quadratic where magnitude of the SOCs are assumed to be equal. Our calculation and the graphical analysis (shown in Figures 3, 4, 5, and 6) with extremely low-lying states are presented in Sec. III. The paper ends with brief discussion and concluding remarks.

**II. Hamiltonian of graphene on TMD substrate**

We consider the detailed Hamiltonian [32,33,34,35] of graphene monolayer on a two dimensional transition metal di-chalcogenide built on the orbital Hamiltonian for pristine graphene. The detailed Hamiltonian is basically a low-energy one around the Dirac points **K** and **K′** in the basis $(a^\xi_{k\uparrow}, b^\xi_{k\downarrow}, a^\xi_{k\downarrow}, b^\xi_{k\uparrow})$ in momentum space. Here $a^\xi_{k\sigma}$ ($b^\xi_{k\sigma}$) is the fermion annihilation operator for the state $(k,\sigma)$ corresponding to the valley $\xi = \pm 1$, and the sub-lattice A(B). The Hamiltonian consists of the sub-lattice-resolved, giant intrinsic spin-orbit couplings(SOC) due to the hybridization of the carbon orbitals with the d-orbitals of W/Mo. These couplings correspond to next-nearest neighbor hopping without spin-flip. There is extrinsic Rashba spin-orbit coupling as well, that allows for external tuning of the band gap in G+TMD and connects the nearest neighbors with spin-flip. It arises because the inversion symmetry is broken when graphene is placed on top of a TMD. There is orbital gap related to the transfer of the electronic charge from graphene to TMD. The all four substrate-induced inter-action terms are time-reversal invariant and absent by inversion symmetry in isolated graphene sheets. Therefore, the low-energy dimensionless Hamiltonian[32,33,34] for a G+TMD system may be written down explicitly as

$$H\left(\tfrac{\hbar v_F}{a}\right) = \sum_{\delta k}\, (a^{\dagger\xi}_{\delta k\uparrow}\ \ b^{\dagger\xi}_{\delta k\downarrow}\ \ a^{\dagger\xi}_{\delta k\downarrow}\ \ b^{\dagger\xi}_{\delta k\uparrow})$$

$$\frac{\mathcal{H}(\delta k)}{\left(\tfrac{\hbar v_F}{a}\right)} \begin{pmatrix} a^\xi_{\delta k\uparrow} \\ b^\xi_{\delta k\downarrow} \\ a^\xi_{\delta k\downarrow} \\ b^\xi_{\delta k\uparrow} \end{pmatrix} \quad (1)$$

$$\frac{\mathcal{H}(\delta k)}{\left(\tfrac{\hbar v_F}{a}\right)} = \begin{pmatrix} a_1 & h^\xi_+ & -(\lambda^+ + \lambda^-)i\delta k_- & -a\delta k^\xi_+ \\ h^{\xi*}_+ & a_2 & -a\delta k^\xi_- & -(\lambda^+ - \lambda^-)i\delta k_+ \\ (\lambda^+ + \lambda^-)i\delta k_+ & -a\delta k^\xi_+ & a_3 & h^\xi_- \\ -a\delta k^\xi_- & (\lambda^+ - \lambda^-)i\delta k_- & h^{\xi*}_- & a_4 \end{pmatrix}, (2)$$

$a_1 = \Delta + M + \xi \Delta^A_{soc},\ a_2 = -\Delta - M + \xi \Delta^B_{soc},\ a_3 = \Delta - M - \xi \Delta^A_{soc},$

$$a_4 = -\Delta + M - \xi \Delta^B_{soc}$$

$$h^\xi_+ = \tfrac{3}{2} i\lambda_R(E)(1+\xi),$$
$$h^\xi_- = \tfrac{3}{2} i \lambda_R(E)(1-\xi). \quad (3)$$

Here the nearest neighbor hopping is parameterized by a hybridization $t$, $\left(\tfrac{\hbar v_F}{a}\right) = \left(\tfrac{\sqrt{3}}{2}t\right)$, and a = 2.46 A° is the pristine graphene lattice constant. Also, $\delta k^\xi_\pm \rightarrow \delta k_\pm$ (that is, $\delta k_x \pm i\, \delta k_y$) for the Dirac point **K** ($\xi = +1$) and $\delta k^\xi_\pm \rightarrow \delta k^*_\pm$ (that is, $\delta k_x \mp i\, \delta k_y$) for the Dirac point **K′** ($\xi = -1$). The quantity $E(s_z, t_z) = \xi\, t'_{so}\, s_z\, t_z + \Delta_z\, t_z + M\, s_z$, with the spin index $s_z = \pm 1$ and the sub-lattice pseudo-spin index $t_z = \pm 1$. The parameters orbital proximity gap $\Delta$, the intrinsic parameters $\Delta^A_{soc}$ and $\Delta^B_{soc}$, and the extrinsic Rashba parameter $\lambda_R(E_z)$ allow for tuning by the applied electric field. Since the WSe$_2$/MoS$_2$ layer provides different environment to atoms A and B in the graphene-cell, there is this (dimension-less) orbital proximity gap

$$\Delta = \frac{\Delta_{Orbital}}{\left(\tfrac{\hbar v_F}{a}\right)}$$

arising from the effective staggered potential induced by the pseudo-spin symmetry breaking. The orbital gap $\Delta_{Orbital}$ is about **0.5 meV [32,33,34]** in the absence of electric field. When the field crosses a limiting value 0.5 V/nm, the gap exhibits a sharp increase. This gap is related to the transfer of the electronic charge from graphene to TMDs. The quantum anomalous Hall **s**tate could be accessed in G+TMD by introducing an exchange field. The exchange field $M'$ ($M = M'/\left(\tfrac{\hbar v_F}{a}\right)$) arises due to proximity coupling to a ferro-magnet such as depositing Fe atoms to the graphene surface. This modus operandi to extract the exchange coupling effect has been suggested in the case of graphene and silicene by several authors **[37,38,39,40].** Due to the hybridization of the carbon orbitals with the d-orbitals of W/Mo, there is sub-lattice-resolved, giant intrinsic spin-orbit couplings($t^A_{so}, t^B_{so}$)

$$\Delta^A{}_{soc}= \frac{t^A_{so}}{\left(\frac{\hbar v_F}{a}\right)}, \quad \Delta^B{}_{soc}= \frac{t^B_{so}}{\left(\frac{\hbar v_F}{a}\right)}.$$

**TABLE 1**

| TMDC | T [eV] | $\Delta_{Orbital}$ [meV] | $t^A_{so}$ [meV] | $t^B_{so}$ [meV] | $\lambda'_R$ [meV] |
|---|---|---|---|---|---|
| WSe$_2$ | 2.51 | 0.54 | −1.22 | 1.16 | 0.56 |
| WS$_2$ | 2.66 | 1.31 | −1.02 | 1.21 | 0.36 |
| MoSe$_2$ | 2.53 | 0.44 | −0.19 | 0.16 | 0.26 |
| MoS$_2$ | 2.67 | 0.52 | −0.23 | 0.28 | 0.13 |

These couplings correspond to next-nearest neighbor hopping **[32,33,34]** without spin-flip. The spin-orbit field parameters for **G+TMD** are about **0.50 meV**, which is 20 times more than that in pure graphene **[32,33,34,35]** ($t_{soc}$ ~ 24 μeV). The parameter

$$\lambda_R = \lambda'_R / (\hbar v_F/a)$$

is the extrinsic Rashba spin-orbit coupling (RSOC), that allows for external tuning of the band gap in G+TMD and connects the nearest neighbors with spin-flip. It, thus, arises because the inversion symmetry is broken when graphene is placed on top of a TMD. While the intrinsic parameters $\Delta^A{}_{soc}$ and $\Delta^B{}_{soc}$ change rather moderately with the increase in the applied electric field, the Rashba parameter $\lambda_R$ almost doubles in increasing the field from −2 to 2 V/nm. For the pristine graphene $\lambda'_R \approx$ **10 μeV** whereas for GTMDC(WSe$_2$) it is 0.56 meV. Wang et al.**[35]**, however, have reported it to be approximately 1 meV. The spin-splitting by the Rashba term away from the points **K** and **K′** is the same as that at **K** and **K′**. The three spin-orbit interaction terms, with coupling constant ($t^A_{so}$, $t^B_{so}$) and $\lambda'_R$, are induced by interfacial interactions. The all four substrate-induced interaction terms, $\Delta_{Orbital}$, ($t^A_{so}$, $t^B_{so}$) and $\lambda'_R$, are time-reversal invariant and absent by inversion symmetry in isolated graphene sheets. The sub-lattice resolved, pseudo-spin inversion asymmetry driven spin-orbit coupling term (ASOC), on the other hand, represents the next-nearest-neighbor, unlike the Rashba term, same sub-lattice hopping away from **K** and **K′** albeit with a spin flip. In the basis ($a_{k\uparrow}$, $b_{k\downarrow}$, $a_{k\downarrow}$, $b_{k\uparrow}$), the ASOC terms, involving

$$\lambda^+ = \lambda'^+ / \left(\frac{\hbar v_F}{a}\right), \quad \lambda^- = \lambda'^- / \left(\frac{\hbar v_F}{a}\right),$$

could be written in a manner as shown in Eq.(2). Here $\lambda^{'+}$ and $\lambda^{'-}$, respectively, are the spin-orbit interactions representing the average coupling, and the differential coupling between the A and B sub-lattices. Apart from these parameters, there is the intrinsic, momentum-dependent Rashba interaction which is modeled as $\alpha(\delta k_y \sigma_x - \delta k_x \sigma_y)$ where $\sigma's$ are the Pauli matrices. We have not taken these into account in this preliminary treatment. Some of the values of the orbital and spin-orbital parameters are summarized in table 1. These parameters can be tuned by a transverse electric field and vertical strain. As could be seen in this table, the sum of the absolute value of the intrinsic SOC terms is *greater* than the term $\Delta_{Orbital}$ characterizing the (staggered) sub-lattice asymmetry in the graphene A and B atoms on WSe$_2$ and WS$_2$ whereas it is less for MoY$_2$. It was recently shown by Ulloa et al.**[41]** that as long as the former is valid the anti-crossing of bands with opposite spins takes place, due to the presence of the Rashba term, around each of the valleys near the **K** point of graphene, However, when the latter is true, one makes a cross-over to a 'direct band' regime with typically parabolic dispersion for each of the two spin projections, .

The energy eigen-values ($E(a|\delta k|)$) of the matrix (2) are given by a quartic. In terms of the powers of ε (where $\varepsilon \equiv E(a|\delta k|)/\lambda_R$), in the absence of PIA driven terms, the quartic may be written as $\varepsilon^4 - 2\varepsilon^2 b + b^2 = 4\varepsilon c + b^2 - d$, where

$$a = 0, \quad b_\xi(|\delta k|, M) = \left[\left(\frac{\Delta}{\lambda_R}\right)^2 + \frac{\left(\frac{\Delta^A_{Soc}}{\lambda_R}\right)^2 + \left(\frac{\Delta^B_{Soc}}{\lambda_R}\right)^2}{2} + (9/4)(1+\xi^2) + \left(\frac{M}{\lambda_R}\right)^2 + \left(\frac{\hbar v_F}{a}\right)^2 \frac{(a|\delta k|)^2}{\lambda_R^2} - \xi\left\{\left(\frac{\Delta^B_{Soc}}{\lambda_R}\right) + \left|\left(\frac{\Delta^A_{Soc}}{\lambda_R}\right)\right|\right\}\left(\frac{M}{\lambda_R}\right)\right],$$

$$c_\xi(M) = \xi\left\{\left(\frac{\Delta^B_{Soc}}{\lambda_R}\right) - \left|\left(\frac{\Delta^A_{Soc}}{\lambda_R}\right)\right|\right\}\left[\left(\frac{9}{4}\right) - \left(\frac{\Delta}{2\lambda_R}\right)\xi\left\{\left(\frac{\Delta^B_{Soc}}{\lambda_R}\right) + \left|\left(\frac{\Delta^A_{Soc}}{\lambda_R}\right)\right|\right\} + \left(\frac{\Delta}{\lambda_R}\right)\left(\frac{M}{\lambda_R}\right)\right],$$

$$d_\xi(|\delta k|, M) = \sum_{j=1}^{5} d_j,$$

$$d_1 = \left\{\left(\frac{\Delta}{\lambda_R}\right)^2 - \left(\xi\left(\frac{\Delta^B_{Soc}}{\lambda_R}\right) - \left(\frac{M}{\lambda_R}\right)\right)^2\right\} \times \left\{\left(\frac{\Delta}{\lambda_R}\right)^2 - \left(\xi\left|\left(\frac{\Delta^A_{Soc}}{\lambda_R}\right)\right| - \left(\frac{M}{\lambda_R}\right)\right)^2\right\},$$

$$d_2 = (9/4)(1+\xi)^2\left[\left\{\left(\frac{\Delta}{\lambda_R}\right) - \left(\frac{M}{\lambda_R}\right)\right\}^2 + \xi\left\{\left(\frac{\Delta^B_{Soc}}{\lambda_R}\right) + \left|\left(\frac{\Delta^A_{Soc}}{\lambda_R}\right)\right|\right\}\left\{\left(\frac{\Delta}{\lambda_R}\right) - \left(\frac{M}{\lambda_R}\right)\right\} + \left(\frac{\Delta^B_{Soc}}{\lambda_R}\right)\left|\left(\frac{\Delta^A_{Soc}}{\lambda_R}\right)\right|\right],$$

$$d_3 = (9/4)(1-\xi)^2\left[\left\{\left(\frac{\Delta}{\lambda_R}\right) + \left(\frac{M}{\lambda_R}\right)\right\}^2 - \xi\left\{\left(\frac{\Delta^B_{Soc}}{\lambda_R}\right) + \left|\left(\frac{\Delta^A_{Soc}}{\lambda_R}\right)\right|\right\}\left\{\left(\frac{\Delta}{\lambda_R}\right) + \left(\frac{M}{\lambda_R}\right)\right\} + \left(\frac{\Delta^B_{Soc}}{\lambda_R}\right)\left|\left(\frac{\Delta^A_{Soc}}{\lambda_R}\right)\right|\right],$$

$$d_4 = \left(\frac{\hbar v_F}{a}\right)^2 \left(\frac{2(a|\delta k|)^2}{\lambda_R^2}\right)\left[\left\{\left(\frac{\Delta}{\lambda_R}\right)^2 - \left(\frac{M}{\lambda_R}\right)^2\right\} + \xi\left\{\left(\frac{\Delta^B_{Soc}}{\lambda_R}\right)\right.\right.$$

$$+|\Delta^A_{soc}/\lambda_R|\}(M/\lambda_R) - (|\Delta^A_{soc}/\lambda_R|)(\Delta^B_{soc}/\lambda_R)],$$

$$d_5 = (\hbar v_F/a)^4 ((a|\delta k|)^4/\lambda^4_R) . \qquad (4)$$

We write $b_\xi(a|\delta k|, M) = \varepsilon^2_{\delta k} + \beta^2_\xi(M)$, where $\varepsilon^2_{\delta k} = (\hbar v_F/a)^2 ((a|\delta k|)^2/\lambda^2_R)$, and

$$\beta^2_\xi(M) = [(\Delta/\lambda_R)^2 + (1/2)\{|\Delta^A_{soc}/\lambda_R|^2 + (\Delta^B_{soc}/\lambda_R)^2\} + (M/\lambda_R)^2$$

$$+ (9/4)(1+\xi^2) - \xi\{|\Delta^A_{soc}/\lambda_R| + (\Delta^B_{soc}/\lambda_R)\} (M/\lambda_R)]. \qquad (5)$$

We now add and subtract an as yet unknown variable $z$ within the squared term ($\varepsilon^4 - 2\varepsilon^2 b + b^2$):

$$(\varepsilon^2 - b + z - z)^2 = 4\varepsilon c + b^2 - d \qquad (6)$$

$$(\varepsilon^2 - b + z)^2 = 2z\varepsilon^2 + 4\varepsilon c + (z^2 - 2bz + b^2 - d). (6a)$$

The necessity of retaining the relatively small term ($4\varepsilon c$) in Eq.(6) will be clear towards the end. Upon retaining the term ($4\varepsilon c$), Eq.(6) or, the equation $\varepsilon^4 - 2\varepsilon^2 b - 4\varepsilon c + d = 0$ becomes evidently a quartic whereas ignoring it will give rise to a bi-quadratic with values of '$\varepsilon$' given by $\varepsilon^2 \approx b \pm \sqrt{(b^2 - d)}$. We shall see that without the term ($4\varepsilon c$) an appropriate theoretical discussion of the collective mode of GTMD, which is our future task, does not seem to be possible. The left-hand side of (6) or (6a) is a perfect square in the variable $\varepsilon$. This motivates us to rewrite the right hand side in that form as well. Therefore we require that the discriminant of the quadratic in the variable $\varepsilon$ to be zero. This yields

$$16c^2 - 8z(z^2 - 2bz + b^2 - d) = 0$$

or, $$z^3 - 2bz^2 + (b^2 - d)z - 2c^2 = 0. \qquad (7)$$

The cubic equation above has the discriminant function

$$\aleph = 8b^3c^2 - 72bdc^2 + 4d(b^2-d)^2 - 108c^4. \qquad (8)$$

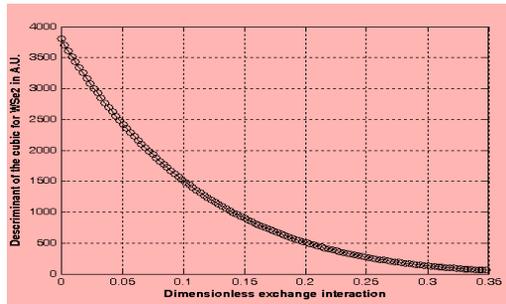

(a)

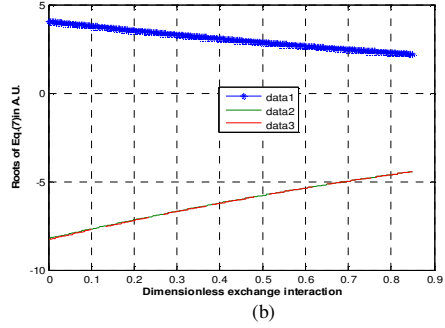

(b)

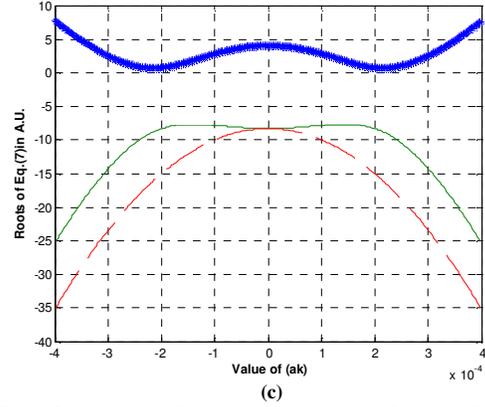

(c)

**Figure 2.** **(a)** A plot of the descriminant $\aleph$ as a function of $M$. **(b)** The plots of the three (real and distinct) roots of (7) as functions of $M$. The uppermost curve (data 1) corresponds to the admissible roots as this is positive. **(c)** The plots of the three (real and distinct) roots of (7) as functions of $(a\delta k)$. The blue line corresponds to $z_0(a\delta k, M)$. We find that $z_0(a\delta k, M) = z_0(-a\delta k, M)$.

Since $\aleph$ is positive as could seen from the figure below (we have plotted here $\aleph$ as a function of '$M$' at a given $(a\delta k) = 0.001$ for graphene on $MoS_2$) we definitely have real roots of Eq.(7). These roots, as functions of '$M$' are shown below in Figure 2. The root corresponding to the uppermost line in Figure 2(b) is the appropriate one as it is found to be real, rational, and, importantly, being of positive sign yields real eigenvalues. Suppose we denote this root by $z_0(a\delta k, M)$. We find that $z_0(a\delta k, M) = z_0(-a\delta k, M)$. Using (6) and (7) one may then write

$$\varepsilon^2 = b - z_0 \pm \{\sqrt{(2z_0)}\varepsilon + \sqrt{(2/z_0)}c\},$$

or,

$$\varepsilon^2 - \sqrt{(2z_0)}\varepsilon + (-b + z_0 - c\sqrt{(2/z_0)}) = 0$$

and

$$\varepsilon^2 + \sqrt{(2z_0)}\varepsilon + (-b + z_0 + c\sqrt{(2/z_0)}) = 0.$$

These two equations basically yield the band structure

$$\varepsilon_{\xi,s,\sigma}(a|\delta k|, M) = [s\sqrt{(z_0/2)} + \sigma\{\varepsilon^2_{\delta k} + \lambda_s(\xi, M)^2\}^{1/2}] , (9)$$

$$\lambda_s(\xi,M)^2 = \{\beta^2_\xi(M) - (z_0/2) + s\sqrt{(2c^2_\xi(M)/z_0)}\}, \quad (10)$$

which consists of two spin chiral conduction bands and two spin chiral valence bands. The bands (in short – hand notation $\varepsilon_{\xi,s,\sigma}(a|\partial k|, M)$ appear as the spin-valley resolved and particle-hole symmetric as $z_0(a\partial k, M) = z_0(-a\partial k, M)$. There is an effective Zeeman field $(s\sqrt{(z_0/2)}$ in Eq.(9)) due to the interplay of the substrate induced interactions with RSOC as the prime player. Because of the spin-mixing driven by the Rashba coupling, the spin is no longer a good quantum number. Therefore the resulting angular momentum eigenstates may be denoted by the spin chirality index $s = \pm 1$. Here $\sigma = + (-)$ indicates the conduction (valence) band. In our analysis (see Figure3), the rele-vant band gap $(G_\xi(a|\partial k|, M))$ for a given valley is found to be the energy difference between the spin-down conduction band and the spin-up valence band: $G_\xi(a|\partial k|, M) = \varepsilon_{\xi,\downarrow,+1}(a|\partial k|, M) - \varepsilon_{\xi,\uparrow,-1}(a|\partial k|, M)$. Furthermore, in Figure 3(left-panel), one notices the anti-crossing of these non-parabolic bands with opposite spins around **K** point for $WSe_2$ and M = 0. As could also be seen in this figure (right-panel), the two bands have a huge band-gap for $WS_2$ with M = 0.25. The other non-parabolic bands, viz. spin-up conduction and spin-down valence, are far-away from the zero energy line.

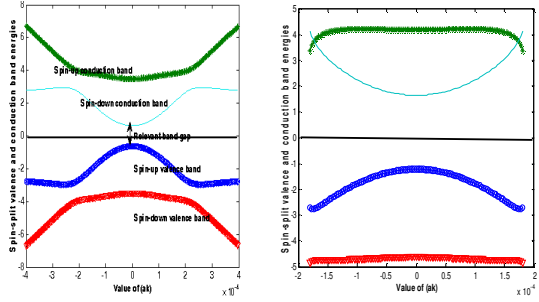

**Figure 3.** The 2-D plots of the spin-split conduction and valence band energies for graphene on $WSe_2$ and $WS_2$ at the Dirac point **K** as a function of the dimensionless wave vector $(a|\partial k|)$. The exchange field equal zero for the left panel ($WSe_2$). For the right panel ($WS_2$) it is 0.25.

**III. Graphical analysis with low-lying states**

In the absence of the substrate-induced interactions(RSOC is present though ) and the exchange interactions, the band structure reduces to the spin-resolved, valley-degenerate energy dispersion of the graphene , viz. $\varepsilon_{\xi s,\sigma}(a|\partial k|, M) = [s\sqrt{(z_0/2)} + \sigma\varepsilon_{\partial k}]$. If the RSOC is absent as well, then the band-structure reduces to the spin-valley degenerate energy dispersion of the pristine, pure graphene: $\grave{e}_\sigma = \sigma\varepsilon_{\partial k}$. It is gratifying to note that all the complexities present in the band structure is woven around the dispersion of the pure graphene. Moving over to Eq.(9), we notice that RSOC $(\sqrt{(z_0/2)}\lambda_R)$ here acts as an in-plane Zeeman term $g_b\mu_B B$ (where B is the Zeeman field, and $g_b$ is the Lande g-factor, and $\mu_B$ is the Bohr magneton). The Zeeman term of the spectrum (9) comes into being due the presence of the term ( 4 $\varepsilon$ c ) in (6). Without the term (4 $\varepsilon$ c ), the spectrum reduces to a bi-quadratic (with no Zeeman term) rather than a quartic. The Zeeman field albeit the Rashba SOC with negligible orbital effects, in conjunction with the spin-orbit coupling (SOC), ushers in the spin-polarization (P =$(n_\uparrow - n_\downarrow)/(n_\uparrow + n_\downarrow)$ where $n_\sigma$ is the spin polarized carrier density in graphene). The ushering in due to the former, i.e. Rashba SOC, could be easily understood by recalling the result for the polarization: P ≈ −sgn(B) $(g_b\mu_B B/(\hbar v_F k_F))^2$ for small magnetic fields. Here $k_F = \sqrt{(\pi n)}$ is the Fermi momentum in un-polarized graphene, where 'n' is the carrier concentration. The ushering in act of the latter, viz. intrinsic SOC, is easy to understand in a qualitative way: We recall that spin-orbit coupling is the natural outcome of incorporating special relativity within quantum mechanics. The external electric field along with that from the atomic cores is Lorentz transformed into an effective magnetic field in the rest frame of an electron moving through a lattice. This effective field, subsequently, acts upon the spin of the electron. It may be mentioned that the spin-orbit interaction utilization for manipulating the electron spin has several distinct advantages, such as, the obviation of the design complexities that are often associated with incorporating local magnetic fields into a device architectures. The polarized electron spins in graphene may be probed through their interaction with optical fields. The polarization of light incident on the graphene will rotate in proportion to the strength of the magnetic field produced by this spin polarization. The rotation is known as the Faraday (Kerr) effect in transmission (reflection). The spin-orbit coupling generates spin polarization through yet another route: the (spin-dependent) skew scattering of relativistic electrons by a Coulomb potential in which electrons with spin up and down are scattered in opposite trajectories **[42].** The extensive investigation of these issues, however, have been relegated to a future communication.

Our calculation and the graphical analysis (shown in Figures 3, 4, 5, and 6) with extremely low-lying states strongly suggests that graphene on $WSe_2$ and $WS_2$ exhibits band non-crossing. The bands are strictly particle-hole symmetric despite retaining the term (4 $\varepsilon$ c ) in Eq.(6). The particle-hole symmetry in this context holds if we linearize the band structure near the Fermi level, so that filled states above the Fermi level and empty states below it have the same dispersion. The property to have the spectrum symmetry $\varepsilon_{\xi,s,\sigma}(a\delta k, M) = \varepsilon_{\xi,s,\sigma}(-a\delta k, M)$ is mandatory to have a particle-hole symmetry in this case. This is what is precisely not lacking in Eq.(9) as $z_0(a\delta k, M) = z_0(-a\delta k, M)$. The meaning used here is different from the particle-hole (or charge conjugation) symmetry property of the mean-field theory of superconductivity where this property corresponds to an anti-unitary operator involving the anti-commutation of the Hamiltonian with the same.

In fact, the graphene on TMD is gapped at all possible exchange field values (see Figure 3) in our problem with plethora of perturbations. On account of the strong spin-orbit coupling, the system acts as a quantum spin Hall insulator for $M = 0$. As the exchange field ($M$) increases, the band gap narrowing takes place followed by its recovery. The essential features of these curves, apart from the particle-hole symmetry, are (i) opening of an orbital gap due to the effective staggered potential, (ii) spin splitting of the bands due to the Rashba spin-orbit coupling and the exchange coupling, and (iii) the band gap narrowing and widening due to the many-body effect and the Moss-Burstein effect **[42,43,44,45]**respectively. The latter is due to the enhanced exchange effect. The exchange field $M$ arises due to proximity coupling to ferromagnetic impurities, such as depositing Fe atoms to the graphene surface. Our plot in Figure 4 for the Dirac point **K** shows that as the exchange field increases the relevant band gap between the spin-down conduction band and the spin-up valence band gets narrower followed by the gap recovery and the gap widening. The contour plot of this band-gap as a func-tion of the dimensionless wave vector and the exchange field ($M$) in the case of graphene on WSe$_2$ corroborates this fact. The plots in Figure 5 show the spin-polarization inversion for values of $M$ away from zero. The darker region in Figure 5(b) corresponds to the negative values of the polarization. The plots refer to the Dirac point **K**. In Figure 6, we have plotted the spin-polarization density corresponding to the Dirac point **K′** as a function of the dimensionless exchange field for $(a|\delta k|) = 0.0000$ in the case of graphene on MoSe$_2$(Blue curve) and MoS$_2$(Green curve) at T = 10 K. The plots also indicate the occurrence of the polarization inversion.

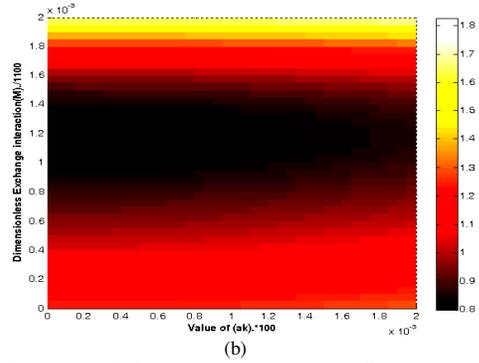

(b)

**Figure 4.** (a)2-D plots of the relevant gap(G) as a function of the dimensionless exchange field for $(a|\delta k|) = 0.0000$ in the case of graphene on WSe$_2$ and on WS$_2$ (In the inset we have plotted the relevant gap as a function of the exchange field for MoSe$_2$ (blue line)and MoS$_2$(green line)). As the exchange field increases the band gap gets narrower followed by the gap recovery and the widening for WSe$_2$ .For WS$_2$ only the gap gets narrower. These patterns are almost replicated by MoSe$_2$ and MoS$_2$. (b)The contour plot of the relevant band-gap as a function of the dimensionless wave vector and the exchange field ($M$) in the case of graphene on WSe$_2$. This plot corroborates that, indeed, the band gap gets narrower followed by the gap recovery with the increase in the exchange field. The plots refer to the Dirac point **K**.

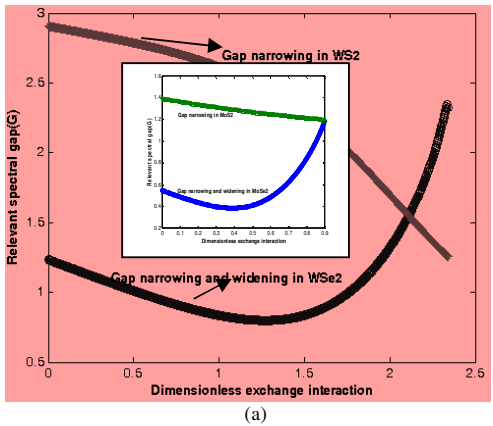

(a)

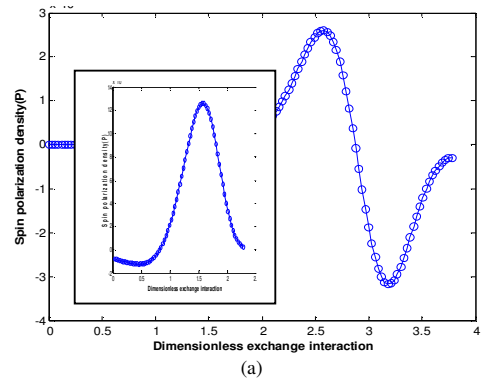

(a)

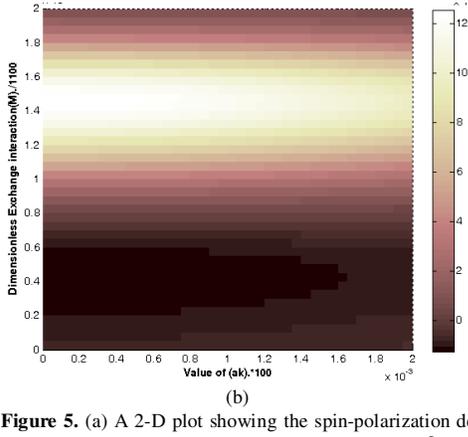
(b)

**Figure 5.** (a) A 2-D plot showing the spin-polarization density as a function of the dimensionless exchange field for $(a|\delta k|) = 0.0000$ in the case of graphene on $WS_2$ at T = 1 K. In the inset we have plotted the spin-polarization density as a function of the exchange field for $WSe_2$. The plots show polarization inversion for values of *M* away from zero. (b) A contour plot showing the spin-polarization density as a function of the dimensionless wave vector and the exchange field (*M*) in the case of graphene on $WSe_2$ at T = 1 K. The plot and the color-bar indicate the occurrence of the polarization inversion. The darker region corresponds to the negative values of the polarization. The plots refer to the Dirac point **K**.

As regards $MoY_2$, we find that there is Moss-Burstein (MB) shift only and no band narrowing. The shift due to the MB effect is usually observed due to the occupation of the higher energy levels in the conduction band from where the electron transition occurs instead of the conduction band minimum. On account of the MB effect, optical band gap is virtually shifted to high energies because of the high carrier density related band filling. This may occur with the elastic strain as well. Thus, studies are required to establish the simultaneous effect of the strain field and the carrier density on optical properties of GTMD. We note that the band gap narrowing and the $v_F$ renormalization, both, in Dirac systems, are essentially many body effects. Our observation of the gap narrowing in graphene on $WSe_2$, thus, supports the hypothesis of $v_F$ renormalization **[46]**. Furthermore, (i) the direct information on the gap narrowing and the $v_F$ renormalization in graphene can be obtained from photoemission, which is a potent probe of many body effects in solids, and,(ii) as already mentioned, new mechanisms for achieving direct electric field control of ferromagnetism are highly desirable in the development of functional magnetic interfaces.

The table above shows that $(\Delta^B_{soc}/\lambda_R)$ is approximately equal to the absolute magnitude of $(\Delta^A_{soc}/\lambda_R)$ in all the cases, viz. those of $MoS_2$, $MoSe_2$, $WS_2$, and $WSe_2$. As a result, in the equation $\varepsilon^4 - 2\varepsilon^2 b - 4\varepsilon c + d = 0$, one may be tempted to ignore the term $(4\varepsilon c) \sim \{(|\Delta^A_{soc}/\lambda_R|) - (\Delta^B_{soc}/\lambda_R)\}\{\ldots\ldots\ldots\}$ compared to the other terms.

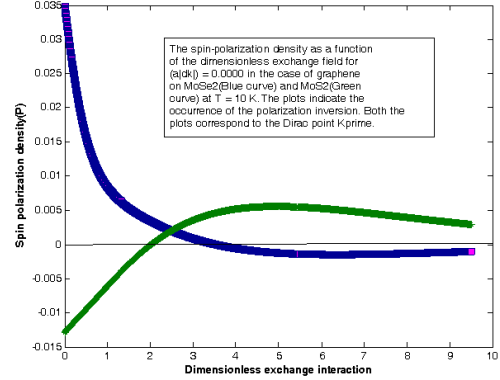

**Figure 6.** The plots of the spin-polarization density as a function of the dimensionless exchange field for $(a|\delta k|) = 0.0000$ in the case of graphene on $MoSe_2$(Blue curve) and $MoS_2$(Green curve) at T = 10 K. The plots indicate the occurrence of the polarization inversion. Both the plots correspond to the Dirac point **K′**.

Thus, we have a bi-quadratic in place of a quartic. This immediately yields $\varepsilon^2 \approx b + s\sqrt{(b^2-d)}$. We shall see below that the approximation is inappropriate as it leads to (i) the non-appearance of the MB shift in the spectral gap between the spin-down conduction band and the spin-up valence band closer to line E= 0, and (ii) the spectra arising out of a bi-quadratic which is inadequate for the discussion of the collective mode in graphene on account of the non-appearance of the crucial Zeeman term to sustain the spin-polarization. We now write the typically particle-hole symmetric band structure $\varepsilon(\delta k,M)$, arising out of $\varepsilon^2 \approx b + s\sqrt{(b^2-d)}$ and which consists of two spin-chiral conduction and two spin-chiral valence bands, as

$$\varepsilon(\delta k,M) = \sigma[\,\mathcal{E}^2_{\delta k} + \Delta^2_{\xi,s}(\delta k,M)]^{1/2}$$

where

$$\mathcal{E}^2_{\delta k} = =(\hbar v_F/a)^2((a|\delta k|)^2/\lambda^2_R),$$

$$\Delta^2_{\xi,s}(\delta k,M) = [(\Delta/\lambda_R)^2 + (1/2)\{|\Delta^A_{soc}/\lambda_R|^2 + (\Delta^B_{soc}/\lambda_R)^2\}$$

$$+ (M/\lambda_R)^2 + (9/4)(1+\xi^2) - \xi\{|\Delta^A_{soc}/\lambda_R| + (\Delta^B_{soc}/\lambda_R)\}(M/\lambda_R)]$$

$$+ s\sqrt{\{b^2_\xi(\delta k,M) - d_\xi(\delta k,M)\}}. \qquad (13)$$

Here we have replaced $(3/2)\lambda_R(E)$ in Eq.(3) by a scaled-down RSOC, viz. $\lambda_R(E)$ and divided rest of the terms in the band structure by this re-defined RSOC. The band index $\sigma = \pm 1$. We remark that the particle-hole symmetry in (13) is totally unaffected by the dropping of the term $(4\varepsilon c)$ compared to the other terms in the equation $\varepsilon^4 - 2\varepsilon^2 b - 4\varepsilon c + d = 0$. Quite interestingly, the spin-polarization does not become zero in this case ( as could be seen in Figure 7) as ignoring the term $(4\varepsilon c)$ simply means the magnitude of the sub-lattice-resolved SOCs are nearly equal. We have shown in Figure 7(a) the plot of the spin-polarization as a function of the dimensionless exchange field(*M*) for (a|δk|) = 0.0000 in the case of graphene on WSe$_2$ at T = 10 K using the dispersion given by (13). The polarization inversion is found to occur at *M* = 0.33. We have also plotted the relevant band gap $G_\xi(a|\delta k|, M) = \varepsilon_{\xi,\downarrow,+1}(a|\delta k|, M) - \varepsilon_{\xi,\uparrow,-1}(a|\delta k|, M)$ in Figure 7 (b) as a function of *M* using Eq.(13). The MB-effect is conspicuous by its absence here.

It is now easy to see that when only this Rashba coupling and the exchange field are present, the band structure is given by the expression

$$\varepsilon_{\xi s,\sigma}(\delta k, M) = \sigma[\varepsilon^2_{\delta k} + +(M/\lambda_R)^2 + 2$$

$$+2s\sqrt{\{1+\varepsilon^2_{\delta k}(1+(M/\lambda_R)^2)\}}]^{1/2}. \quad (14)$$

Equation (14) is the same as the spectrum obtained by MacDonald et al.**[37,47]**. The effective Zeeman field aspect in Eq.(9) is conspicuous by its absence in (14) as the term (4 *ε c*) in Eq.(6) was ignored in the quartic in *ε* to obtain a bi-quadratic. Incidentally, one notices the anti-crossing of the non-parabolic bands in Eq.(14) with opposite spins around **K** point for MoS$_2$ (see Figure 8) and other TMDs due to retaining the Rashba spin-orbit coupling and the exchange field only. The sub-lattice resolved spin-orbit coupling and the effective staggered potential term induced by the pseudo-spin symmetry breaking have seemingly no role to play.

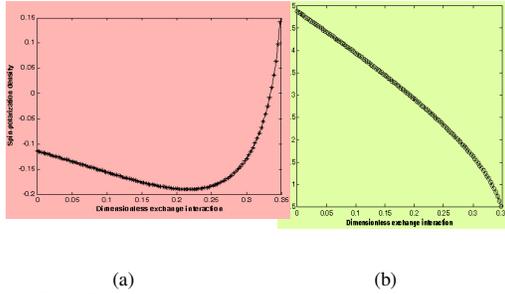

(a)            (b)

**Figure 7.** (a)The plot of the spin-polarization density as a function of the dimensionless exchange field(*M*) for (a|δk|) = 0.0000 in the case of graphene on WSe$_2$ at T = 10 K using the dispersion given by (13). The polarization inversion occurs at *M* = 0.33. (b) The plot of the relevant band gap as a function of *M* using (13). The MB-effect is conspicuous by its absence here.

## IV. Discussion

We once again emphasize that the Zeeman term of the spectrum (9), appearing due to the interplay of the substrate induced interactions with the prime player as the Rashba SOC, is basically due the presence of the term (4 *ε c*) in (6) from the analytical view-point. The Zeeman field albeit the Rashba SOC, in conjunction with the intrinsic spin-orbit coupling (SOC), ushers in the spin-polarization. At this point, digressing slightly, we state our intention is to examine in future the vacuum polarization due to the presence of the electromagnetic field around planar charged fermions in graphene where the space between the fermions, according to the quantum field theory, are inhabited by "virtual" particle–antiparticle pairs fated to get annihilated in the time specified by the energy-time uncertainty relation. Our starting point will be the spectrum given by Eq. (9). These pairs, acting as the dipoles, position themselves in such a way that they partially weaken the field. Evidently, the induced charge density, giving rise to induced potential, correspond to a dielectric effect. The many-body effect here changes the dielectric constant of the sample considerably even at the level of the one-loop contribution. In the same context, we note that a discussion of the stable collective mode of graphene (which corresponds to charge plasmons) requires calculation of the Lindhard function with the inclusion of the spin polarization (or the Zeeman field as we have seen above). The Lindhard dielectric function **[48,49],** which basically captures inter-band transitions due to the electromagnetic field, is dependent both on frequency and momentum. For graphene, with a cavalcade of substrate driven interactions as in the present problem, the Lindhard function is yielded via the random-phase approximation (RPA) including the spin-polarization as a necessary ingredient. Now going back to the issue of the term ($4 \varepsilon c$) in Eq.(6), we observe that without this term the spectral equation (6) obtained by us reduces to a bi-quadratic with no Zeeman term rather than a quartic. The term ($4 \varepsilon c$) eventually led to the Zeeman field in the spectrum (9). It is now clear that any spectrum, such as those given by Eqs.(13) and (14),which arises out of a bi-quadratic is somewhat inadequate for the discussion of the collective mode in graphene due to the absence of the Zeeman term. We remark that the above-mentioned fact is a strong enough reason for proceeding with a quartic as we did in Eq.(6).

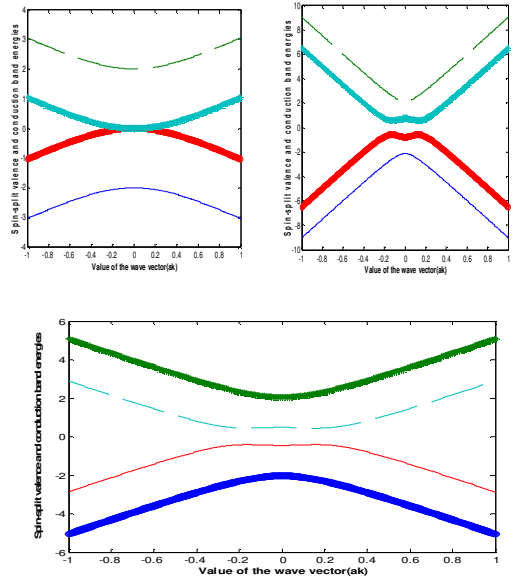

**Figure 8.** A typical band structure corresponding to a spin split semi-metallic phase of the MacDonald et al. model in (14) near the K valley for graphene on WSe$_2$ and M= 0 is in the left panel. In the right panel, we have the anti-crossing of the non-parabolic bands in Eq.(14) with opposite spins around **K** point for graphene on MoS$_2$ and M= 0.10. In the panel below, we have band insulator regime where a finite bulk gap develops for MoSe$_2$ for M= 0.12.

## V. Conclusion

In conclusion, it may be mentioned that the electrical control of magnetic properties is an important research goal for low-power write operations in spintronic data storage and logic[50]. The tuning of the exchange field requires similar kind of the electrical manipulation of magnetism and magnetic properties in a potential experimental observation of the present effect. In the case of thin films of ferromagnetic semiconductors/ insulators, the application of an electric field alters the carrier density which in turn affects the magnetic exchange interaction and the magnetic anisotropy. As already mentioned, the magnetic multi-ferroics, such as BFO, have created quite a stir amongst material research community with the promise of the coupling between the magnetic and electric order parameters. The deeper exploration of this coupling needs to be carried out to have access to electrical control of magnetism through the exchange interaction with a ferromagnet. Finally, we believe our results highlight a promising direction for band gap engineering of graphene.